\documentclass[%
 aip,
 amsmath,amssymb,
 reprint,%
]{revtex4-1}

\usepackage{graphicx}% Include figure files
\usepackage{dcolumn}% Align table columns on decimal point
\usepackage{bm}% bold math
\usepackage[utf8]{inputenc}
\usepackage[T1]{fontenc}
\usepackage{mathptmx}
\usepackage{etoolbox}

\usepackage{xcolor}%

\usepackage{soul} % SKA
\usepackage{float} % SKA
\makeatletter
\def\@email#1#2{%
 \endgroup
 \patchcmd{\titleblock@produce}
  {\frontmatter@RRAPformat}
  {\frontmatter@RRAPformat{\produce@RRAP{*#1\href{mailto:#2}{#2}}}\frontmatter@RRAPformat}
  {}{}
}%
\makeatother

\begin{document}

%\preprint{AIP/123-QED}

%\title[Sample title]{Sample Title:\\with Forced Linebreak}
\title{Optical levitation of fluorescent silicon carbide nanoparticles in vacuum}
% Force line breaks with \\

\author{Seyed Khalil Alavi}
\email{skalavi@fmq.uni-stuttgart.de}
\affiliation{
Institute for Functional Matter and Quantum Technologies, University of Stuttgart, 70569, Stuttgart,
Germany}
\affiliation{
Center for Integrated Quantum Science and Technology (IQST), University of Stuttgart, 70569, Stuttgart,
Germany}

\author{Cheng-I Ho}
\affiliation{
3rd Institute of Physics, University of Stuttgart, 70569, Stuttgart, Germany}
\affiliation{
Center for Integrated Quantum Science and Technology (IQST), University of Stuttgart, 70569, Stuttgart,
Germany}

\author{Iuliia Neumann}
\affiliation{
Institute of Inorganic Chemistry, University of Stuttgart, 70569, Stuttgart, Germany}

\author{Daniel Eberle}
\affiliation{
Institute for Functional Matter and Quantum Technologies, University of Stuttgart, 70569, Stuttgart,
Germany}
\affiliation{
Center for Integrated Quantum Science and Technology (IQST), University of Stuttgart, 70569, Stuttgart,
Germany}

\author{Vadim Vorobyov}
\affiliation{
3rd Institute of Physics, University of Stuttgart, 70569, Stuttgart, Germany}
\affiliation{
Center for Integrated Quantum Science and Technology (IQST), University of Stuttgart, 70569, Stuttgart,
Germany}

\author{Bertold Rasche}
\affiliation{
Institute of Inorganic Chemistry, University of Stuttgart, 70569, Stuttgart, Germany}

\author{Sungkun Hong}
\email{sungkun.hong@fmq.uni-stuttgart.de}
\affiliation{
Institute for Functional Matter and Quantum Technologies, University of Stuttgart, 70569, Stuttgart,
Germany}
\affiliation{
Center for Integrated Quantum Science and Technology (IQST), University of Stuttgart, 70569, Stuttgart,
Germany}

\date{\today}% It is always \today, today,
%\vspace{1cm}
\begin{abstract}

Levitated optomechanics is an emerging field in quantum science that explores the quantum motion of mesoscopic particles levitated in a vacuum. Expanding this approach to particles with intrinsic quantum defects opens new opportunities for quantum sensing and nontrivial quantum state generation. Here, we explore silicon carbide (SiC) nanoparticles as a promising platform that offers a range of controllable quantum defects and material tunability. We demonstrate stable optical levitation of 3C-polytype SiC nanoparticles containing single photon emitters in a vacuum. We observe stable fluorescence from the levitated particle, confirming the preservation of the emitters in the levitated state. We also investigate particle loss at low pressure and explore thermal annealing as a potential method to improve trapping stability. Our results establish SiC as a viable platform for levitated optomechanics, providing additional quantum degrees of freedom and material engineering capabilities.
%Our results establish SiC as a viable material platform for levitated optomechanics with additional quantum degrees of freedom.

\end{abstract}

\maketitle

%Introduction: three paragraph
\section{Introduction}

Nanoscale particles levitated in a vacuum have emerged as a powerful tool for exploring quantum phenomena at the mesoscopic scale \cite{millen_review_2020, gonzalez-ballestero_levitodynamics_2021}. By eliminating mechanical clamping losses inherent to traditional optomechanical systems, vacuum levitation allows for achieving exceptional mechanical quality factors and quantum coherence even at room temperature \cite{chang_cavity_2010, romero-isart_toward_2010}, offering a unique advantage for quantum control experiments. Combined with precise optical measurement and control, the field of levitated optomechanics has enabled remarkable achievements in recent years, including ground-state cooling through both active feedback techniques\cite{magrini_real-time_2021, tebbenjohanns_quantum_2021} and passive cavity-enhanced cooling schemes\cite{delic_cooling_2020, dania_cooling_deep_2024}. These advances in controlling the motion of levitated particles at the quantum level not only provide insights into fundamental quantum mechanics\cite{romero-isart_large_2011, vedral_gravity_entanglement_2017, bose_gravity_entanglement_2017} but also open new possibilities for ultra-sensitive sensors\cite{ranjit_zeptonewton_2016, monteiro_force_2020, ahn_ultrasensitive_2020}.

An intriguing direction in levitated optomechanics explores particles with internal quantum degrees of freedom, for instance, diamond nanocrystals hosting nitrogen-vacancy (NV) centers\cite{neukirch_nv_levitation_2013}. These atomic-like defects possess well-isolated spin states that can potentially couple to the particle's mechanical motion, offering a route to generate non-classical states of motion such as largely delocalized superpositions\cite{yin_large_2013, scala_matter-wave_2013}. Initial demonstrations of optically levitated nanodiamonds with NV centers revealed challenges, particularly particle loss in the crucial medium-vacuum regime\cite{neukirch_nv_levitation_vacuum_2015, tongcang_nv_levitation_vacuum_2016}. This limitation was later addressed through the development of electrical (Paul) trapping schemes, leading to significant breakthroughs, including coherent spin manipulation of a nanodiamond levitated at high vacuum\cite{tongcang_nv_levitation_highvacuum_2024} and spin-assisted cooling of the diamond particle's center-of-mass motion\cite{hetet_spin_cooling_2020}.

Here, we investigate optically levitated silicon carbide (SiC) nanoparticles as a new system for levitated optomechanics with internal quantum degrees of freedom. SiC has gained significant attention in the past decade as a promising material for quantum technologies, hosting a variety of addressable quantum defects\cite{awschalom_defect_quantum_review_nat_photonics_2018,castelletto_sic_quantum_review_2020}. They include optically addressable spin defects with their spin coherence times comparable to NV centers in diamond\cite{awschalom_SiC_long_spin_coherence_nat_mat_2015, wrachtrup_SiC_long_spin_coherence_nat_mat_2015}. A key advantage of SiC lies in its well-established material engineering and fabrication processes\cite{awschalom_defect_quantum_review_nat_photonics_2018,castelletto_sic_quantum_review_2020}, which enable tailoring of the physical shape and material properties of levitated particles, offering new degrees of control and tunability.

In this work, we employed commercially available 3C-polytype SiC nanoparticles and successfully achieved optical levitation of individual particles in a vacuum. We observed stable fluorescence emission from the trapped particles, confirming that the key property of the single photon emitters in the particle remains intact in the levitated state. We found that particle loss occurs at a pressure range similar to previously studied nanodiamonds\cite{neukirch_nv_levitation_vacuum_2015, tongcang_nv_levitation_vacuum_2016}. To mitigate this limitation, we explored thermal annealing to induce partial surface oxidation, leading to moderate improvements in trapping stability. Our work paves the way for future research directions in levitated optomechanics based on SiC and its quantum defects.
\begin{figure*}
\includegraphics[scale=1]{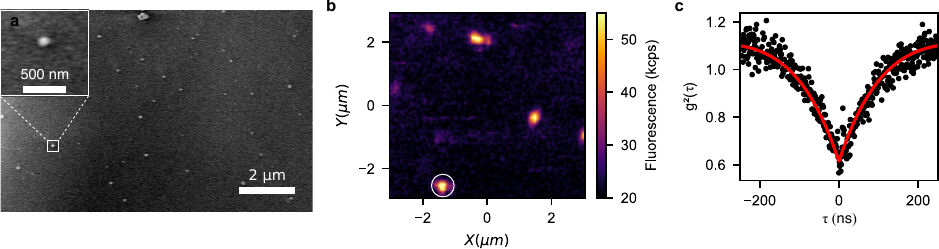}% Here is how to import EPS art
\caption{(a) Scanning electron micrograph of SiC NPs prepared on a glass substrate. The inset in the lower left corner shows an isolated SiC NP with a size of about 100 nm. (b) Confocal scanning microscope image of the similarly prepared SiC particles on a glass substrate, excited by a 514 nm laser. (c) Second-order intensity correlation $g^{(2)}(\tau)$ measured from the fluorescence spot indicated in (b) by a circle.
%automatically numbered
%\vspace{1cm}.
}
\label{fig_char}
\end{figure*}
%
%%%%%%%%%%%%%%%%%%%%%%%%%%%%%%%%%%%%%%%%%
%
\section{Results}
\begin{figure*}[th!]
\includegraphics[scale=1]{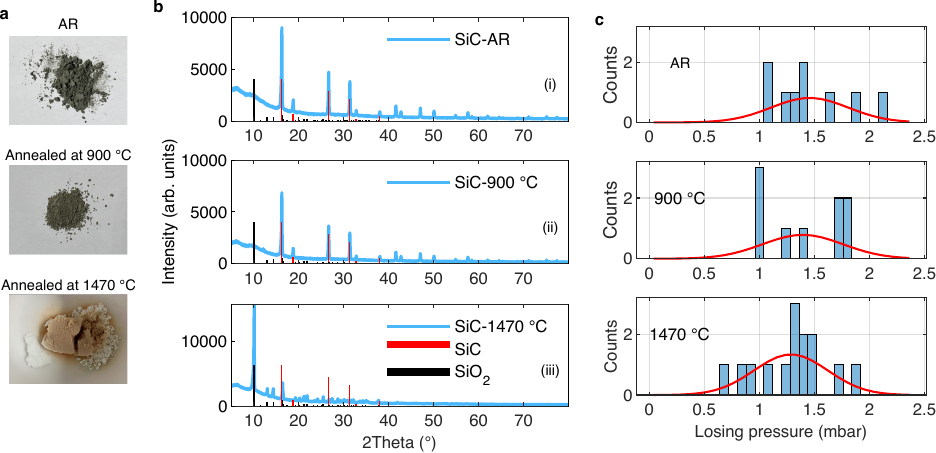}% Here is how to import EPS art
\caption{Effects of annealing on the particle's loss pressure. (a) Optical images of the SiC NP powders as received (AR), after annealing at 900°C for 24 hours (900°C), and after annealing at 1470°C for 24 hours (1470°C). (b) X-ray diffraction (XRD) analyses of as-received SiC NP sample (i), the sample annealed at 900 °C for 24 hours (ii), and 1470 °C for 24 hours (iii). The theoretical patterns for SiC (red) and SiO$_2$ (black) are displayed for reference. All peaks of the as-received and 900 °C-annealed samples can be indexed with SiC. The 1470 °C-annealed sample (iii), in contrast, can be indexed as SiO$_2$. (c) Distribution of measured loss pressures for NP samples prepared under different annealing conditions. The NPs from the as-received sample were lost at a pressure above 1 mbar. The NPs annealed at 1470°C, in contrast, exhibited loss at lower pressure, with some being lost at a pressure below 1 mbar.}
\label{fig_ann}
\end{figure*}

SiC nanoparticles (NP) used in our experiment are commercially available 3C-SiC nanocrystal powders (vendor: Nanoarmor; nominal size: 80-100 nm). 
This sample has previously been shown to host carbon-antisite vacancy (CAV) pairs, which act as bright single-photon emitters \cite{castelletto_SiC_NP_fluorescence_2014}.
We confirmed that the NPs exhibited the specified size distribution (see Fig.\ref{fig_char}a) and also showed characteristic peaks corresponding to SiC from powder X-ray diffraction (PXRD) analysis (see Fig. \ref{fig_ann}b). 

Castelletto et al. \cite{castelletto_SiC_NP_fluorescence_2014} also demonstrated that controlled annealing enables NPs to exhibit stable fluorescence. We followed similar steps to anneal the as-received sample for 3 hours at 900~\textdegree C in air. The detailed procedure is described in supplementary material (SM). We characterized the oxidized sample under a home-built confocal microscope (see Fig. S1 in SM) and observed stable fluorescence under 514 nm laser illumination (see Fig. \ref{fig_char}b). The majority of the measured spots showed the second-order intensity correlation $g^{(2)}(\tau)<1$ at the time delay $\tau~=0$ (see Fig. \ref{fig_char}c, and Fig. S2), confirming that the fluorescence originates from one or a few single-photon emitters.
We speculate that the single-photon emitters we observed are also CAV pair centers, as identified by Castelletto et al.\cite{castelletto_SiC_NP_fluorescence_2014}, since we used the same commercial product and observed a fluorescence emission spectrum in the near-infrared range (see Fig. S3 in SM), similar to the results reported in their study.

%%%%%%%%%%%%%%%%%%%%%%%%%%%%%%%%%%%%%%%%%%%%%%%%%%

% figure 2
\begin{figure}[h]
\includegraphics[scale=1]{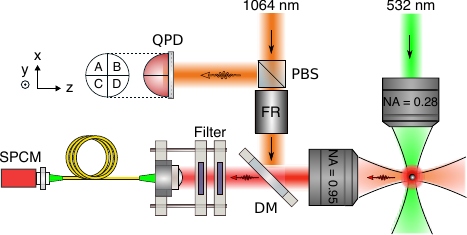}% Here is how to import EPS art
\caption{A schematic representation of the experimental setup combining an optical tweezer with a fluorescence measurement setup. 
The tweezer beam (1064 nm) is sent through the Faraday rotator (FR) and focused by the high NA objective (NA = 0.95). The light scattered by the trapped particle is collected and redirected to a quadrant photodiode (QPD) by FR and a polarization beam splitter (PBS). The trapped particle is excited by a 532 nm beam focused with a 0.28 NA objective. The emitted fluorescence is collected by the tweezer objective, transmitted through a dichroic mirror (DM), and detected via a single-photon counting module (SPCM). Optical filters are used to block any leakage from the tweezer and green excitation light.
}
\label{fig_setup}
\end{figure}
%
%%%%%%%%%%%%%%%%%%%%%%%%%%%%%%%%%%%%%%%%%%%%%%%
%
Next, we used an optical tweezer to levitate a single SiC nanoparticle (NP) in a vacuum. The experimental setup is schematically illustrated in Figure \ref{fig_setup}. The optical tweezer was formed by focusing an intense infrared laser (Azurlight Systems, 1064 nm) through an objective lens with a high numerical aperture (NA = 0.95). The SiC NPs, annealed as previously described, were dispersed in ethanol. This dispersion was then aerosolized and introduced into the chamber under near-ambient conditions. After successfully trapping a single NP, the chamber was evacuated to remove residual particles in it.

\begin{figure}[h]
\includegraphics[scale=1]{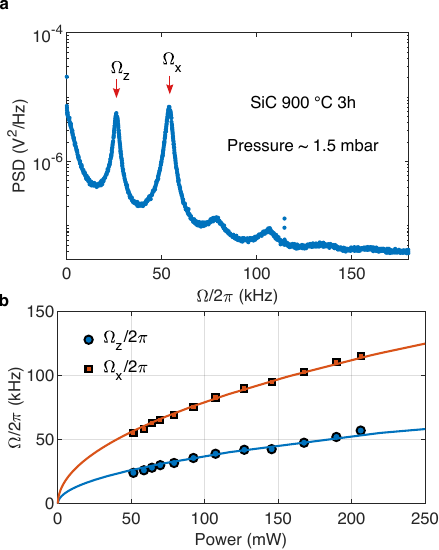}% Here is how to import EPS art
\caption{Optical levitation of a SiC NP (a) Power spectral density of one of the differential signals from the QPD at a pressure of $\sim 1.5~mbar$. The particle's motion along the tweezer beam axis (z-axis) appears as a pronounced peak at the frequency $\Omega_z/2\pi= 26~kHz$. A second peak at $\Omega_x/2\pi= 54~kHz$ corresponds to motion along the x-axis. Motion along the y-axis is not observed, as the differential channel of the QPD used here is not sensitive to it. (b) the particle's frequencies as a function of the tweezer beam power. The frequencies show a square root dependence on the tweezer beam power (solid curves).}
\label{fig_lev}
\end{figure}

%[Levitation: PSD, power vs frequency]
The motion of the trapped NP was monitored by collecting back-scattered light with a tweezer objective and directing it to a quadrant photodiode (QPD) using a Faraday rotator (FR) and a polarizing beam splitter (PBS) placed in the input beam path (see Fig.\ref{fig_setup}). Figure \ref{fig_lev}a presents the power spectral density (PSD) of the signal measured at a pressure of $\sim 1.5~mbar$. The frequency components corresponding to the particle's motion along the tweezer beam's propagation axis and its perpendicular plane were observed.
We further characterized the trapping stability by varying the tweezer beam power, confirming stable particle levitation up to an input laser power of $206~mW$ (Figure \ref{fig_lev}b).

We then illuminated the trapped NP with a green laser ($532~\text{nm}$) to excite fluorescent defects in the NP. This was done by focusing the excitation laser on the trapped NP through a low NA objective (NA = 0.28) aligned perpendicularly to the tweezer objective (Fig. \ref{fig_setup}). The resulting fluorescence emission was collected by the tweezer objective and guided to a single photon counting module (SPCM). The optical path incorporates multiple optical filters including: neutral density ($> 1050~nm$), notch ($532~nm$), and edge ($633~nm$) filters; and a short-pass dichroic mirror (cut-on $900~nm$) to reject the reflected infrared tweezer beam and the green excitation laser, ensuring that only the fluorescent photons (600–800 nm) reach the SPCM.

\begin{figure}%[t!]
\includegraphics[scale=1]{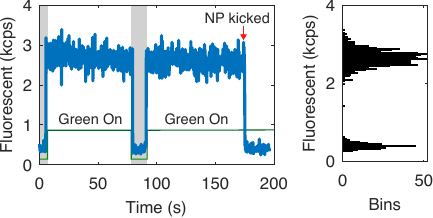}% Here is how to import EPS art
\caption{Fluorescence from the levitated SiC NP measured by SPCM. A stable photon counts of $2652 \pm 203$ counts per second (CPS)  are observed when the green excitation laser is on. When the green laser is off, the photon counts drop down to $403 \pm 71$ CPS.}
\label{fig_fluor}
\end{figure}

Figure \ref{fig_fluor} displays the time trace of photon counts measured by the single-photon counting module (SPCM) at a pressure of $\sim 80~mbar$. The consistent photon count level provides evidence of stable fluorescence emission from the trapped nanoparticle. To verify that these detected photons originate from fluorescence rather than leaked infrared tweezer laser light, we temporarily blocked the green excitation laser. During this interruption, the photon counts dropped to background levels. Upon reintroducing the green excitation beam, the photon counts returned to their previous levels. This confirms that the origin of the counts was the fluorescence from the NP. We note that a majority of the trapped particles exhibited stable fluorescence, some after a brief period of bleaching, which is consistent with observations from our pre-characterization studies of the annealed SiC NPs.

\begin{figure}%[t!]
\includegraphics[scale=1]{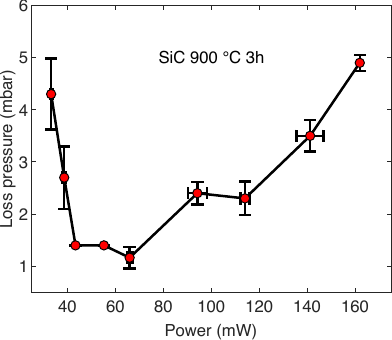}% Here is how to import EPS art
\caption{The loss pressure as a function of tweezer beam power. The observed loss pressure reaches its minimum value of $0.79~mbar$ at the beam power of $67.4~mW$.}
\label{fig_loss}
\end{figure}

Next, we investigated the stability of the NP trapping against the pressure. To that end, we systematically varied the tweezer beam power and measured the pressure at which the particle was lost (Fig. \ref{fig_loss}). 
We observed that the loss pressure decreases with increasing the tweezer power until it reaches the minimum at an input power of $67.4~mW$, above which the loss pressure increases again.
We interpret this behavior as the result of two competing mechanisms. Initially, the loss pressure decreases with increasing optical power, as the depth of the optical potential increases. However, at higher optical powers, the loss pressure begins to rise again as severe optical absorption by the particle destabilizes the motion of the particle. 
Among the factors contributing to absorption are not only the intrinsic bulk absorption properties of SiC, but also optically absorbing defects arising from impurities and structural dislocations.
Nanodiamonds have also been shown in previous studies to exhibit similar absorption-induced loss mechanism \cite{neukirch_nv_levitation_vacuum_2015, tongcang_nv_levitation_vacuum_2016}. 
% 

%%%%%%%%%%%%%%%%%%%%%%%%%%%%%%
% fig

%%%%%%%%%%%%%%%%%%%%%%%%%%%%%%%%%%%%%%%%%%%
% fig 6
%%%%%%%%%%%%%%%%%%%%%%%%%%%%

%
We explored a strategy to reduce the loss pressure by annealing and oxidizing the SiC NPs. Our approach was motivated by two assumptions. First, thermal annealing could potentially heal lattice defects in the crystal structure that contribute to optical absorption. Second, controlled oxidation of SiC can chemically convert the NP into SiO$_{2}$, starting from its surface\cite{alekseev_sic_np_oxidation_2017}, which could provide enhanced structural stability as well as significantly reduced absorption.
We prepared three distinct samples: as-received particles, particles annealed at 900 °C for 24 hours, and particles annealed at 1470 °C for 24 hours (Fig. \ref{fig_ann}a). Visual inspection revealed a progressive color change from dark gray to white with increasing annealing temperature, which is in good agreement with the previous study\cite{alekseev_sic_np_oxidation_2017}. This suggests an increase in the thickness of the oxide layer. 
The increasing oxidation of the SiC-NPs was also confirmed by PXRD (Fig.  \ref{fig_ann}b); while the starting sample and the one annealed at 900 $^\circ C$ can be indexed as SiC, the sample annealed at 1470 $^\circ C$ shows predominantly SiO$_2$ peaks. We note that PXRD is a bulk analysis method and thus is not sensitive to detecting very thin layers of SiO$_2$ (e.g., at 900 $^\circ C$) or very small cores of SiC (e.g., at 1470 $^\circ C$). To that end, we also performed thermal analysis under a synthetic air atmosphere and obtained the thermo-gravimetry (TG) curve (see Fig. S5 in SM). It showed a significant weight uptake indicative of oxidation at around 1000 $^\circ C$. At 1450 $^\circ C$, the curve reaches a maximum at 140 $\%$ of the initial mass. Considering the molar weights of SiC and SiO$_2$, we can derive that 80 mol$-\%$ of the NPs are oxidised to SiO$_2$ and 20 mol$-\%$ remain as SiC, resulting in an approximate core-diameter of 52.6 nm. A detailed procedure for determining the compositional change of the NP from the TG measurements is provided in SM.
This led to a significant reduction in the likelihood of observing stable fluorescence from samples annealed at 1470 $^\circ C$. This is expected, considering that the volume of the SiC was reduced by a factor of 5 during annealing at 1470 $^\circ C$. Prolonged annealing and the resulting healing of lattice defects may also have resulted in an additional reduction in the number of CAV centers.
We additionally confirmed this compositional change by measuring the particles' motional frequencies under identical trapping powers (see Fig. S4 and the caption in SM).

We subsequently investigated whether this annealing-induced modification improved the loss pressure threshold. 
Indeed, some improvement was observed in the annealed sample; however, it was marginal, with the loss pressure still limited to a fraction of a millibar. Given that the absorption coefficient of SiO$_2$ at 1064 nm is exceptionally low, this result suggests that the absorption in our nanoparticle samples is dominated by extrinsic factors—particularly impurities—rather than by the intrinsic absorption properties of SiC.
We note that in order to achieve quantum-limited control of the NP, it is required to achieve stable particle levitation at a pressure lower than $10^{-5}$ mbar.

\section{Conclusion}
In summary, we successfully achieved optical levitation of individual 3C-polytype  silicon carbide (SiC) nanoparticles in a vacuum and demonstrated stable fluorescence emission from the trapped particles. We observed that particle loss occurs at pressures around a few millibars, similar to previous studies on nanodiamonds. To mitigate this limitation, we explored thermal annealing to induce partial surface oxidation, which led to moderate improvements in trapping stability. While annealing did not dramatically improve trapping stability, this highlights a notable advantage of SiC over diamond, i.e., its material tunability. The ability to modify SiC's surface properties opens new possibilities for optimizing its performance and characteristics in various contexts.

The fluorescent defects in SiC used in our experiment, i.e., CAV pair centers, are known to lack addressable spin states\cite{castelletto_sic_quantum_review_2020}. Nevertheless, we emphasize that our study marks an important step for SiC-based levitated optomechanics, as it demonstrated that the defects in SiC remained addressable while the host particle was levitated in a vacuum. In the future, we envisage leveraging SiC’s advantage of facile fabrication to artificially produce high-purity nanoparticles that host defects with controllable spin states. 4H-SiC would be a compelling candidate, as it contains defects with coherent and controllable spin states at room temperature\cite{wrachtrup_SiC_long_spin_coherence_nat_mat_2015}. This will pave the way for future research in levitated optomechanics based on spin–mechanical coupling\cite{yin_large_2013, bose_gravity_entanglement_2017}.

\section*{Supplementary Material}
See the supplementary material for a detailed description of the confocal setup used to characterize the SiC samples and additional analysis of the annealed SiC samples (in particular, the change in NP trapping frequency and the TG measurement of the annealed samples).

\begin{acknowledgments}
This research was supported by the Carl-Zeiss-Stiftung (Johannes-Kepler Grant, IQST) and Vector-Stiftung (MINT-Innovation, Projektantrag: P2023-0100). B.R. and I.N. acknowledge support from the Fond der Chemischen Industrie.
\end{acknowledgments}

%\section*{Author Declarations}
%\subsection*{Conflict of Interest}
%The authors have no conflicts to disclose.

%\subsection*{Author Contributions}

%\section*{Data Availability}
%The data that support the findings of this study are available from the corresponding author upon reasonable request.

%\nocite{*}
%\clearpage
\section*{References}
%\bibliography{aipsamp}% Produces the bibliography via BibTeX.
\bibliography{main_bib}% Produces the bibliography via BibTeX.

%\newpage
%\includepdf[pages=-]{SiC_Optical_levitation_SM.pdf}
\begin{figure*}
\hspace*{-2cm}
    \includegraphics[page = 1, scale =1]{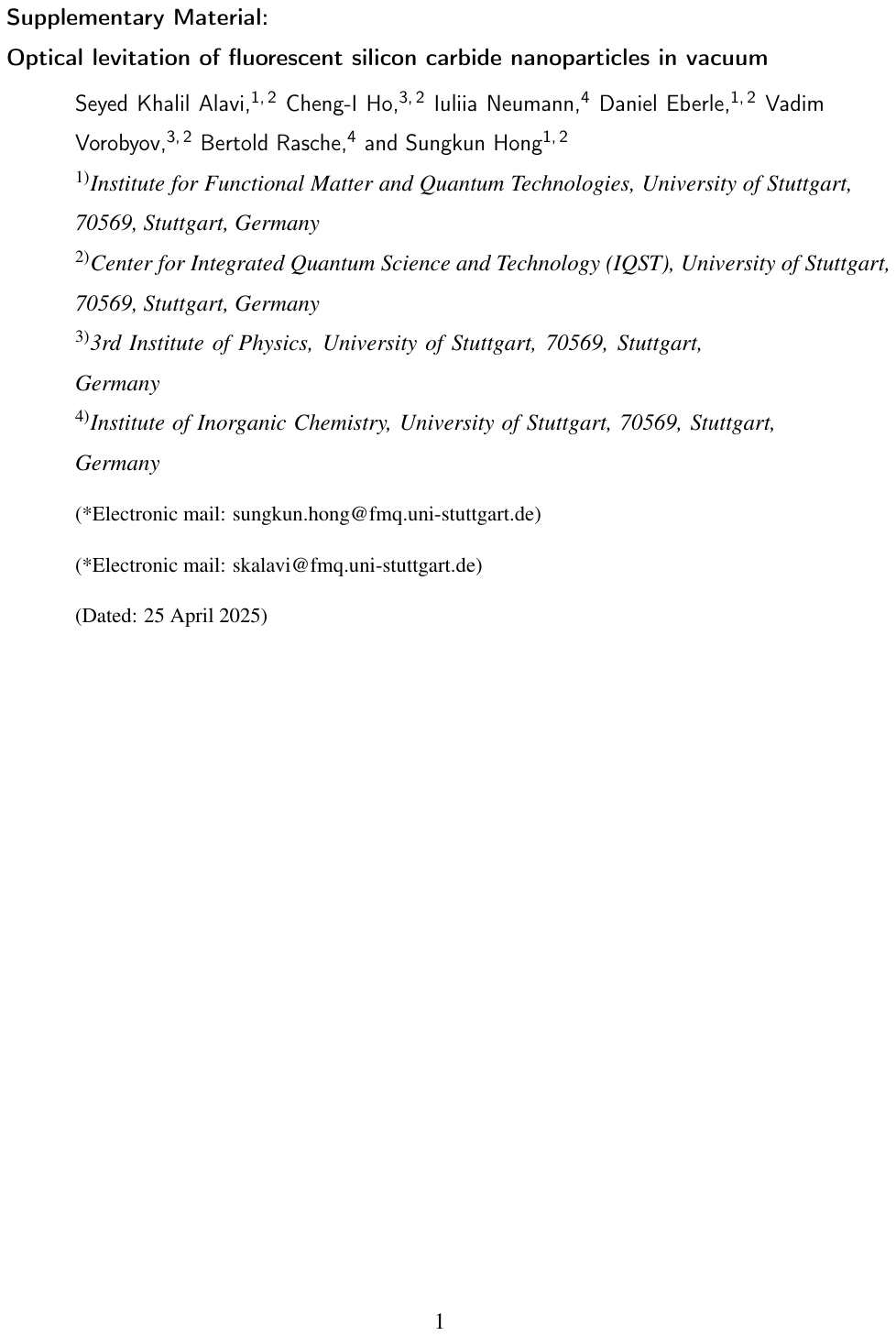}
   \end{figure*}

\begin{figure*}
\vspace*{-1cm}
\hspace*{-2cm}
    \includegraphics[page = 2, scale =1]{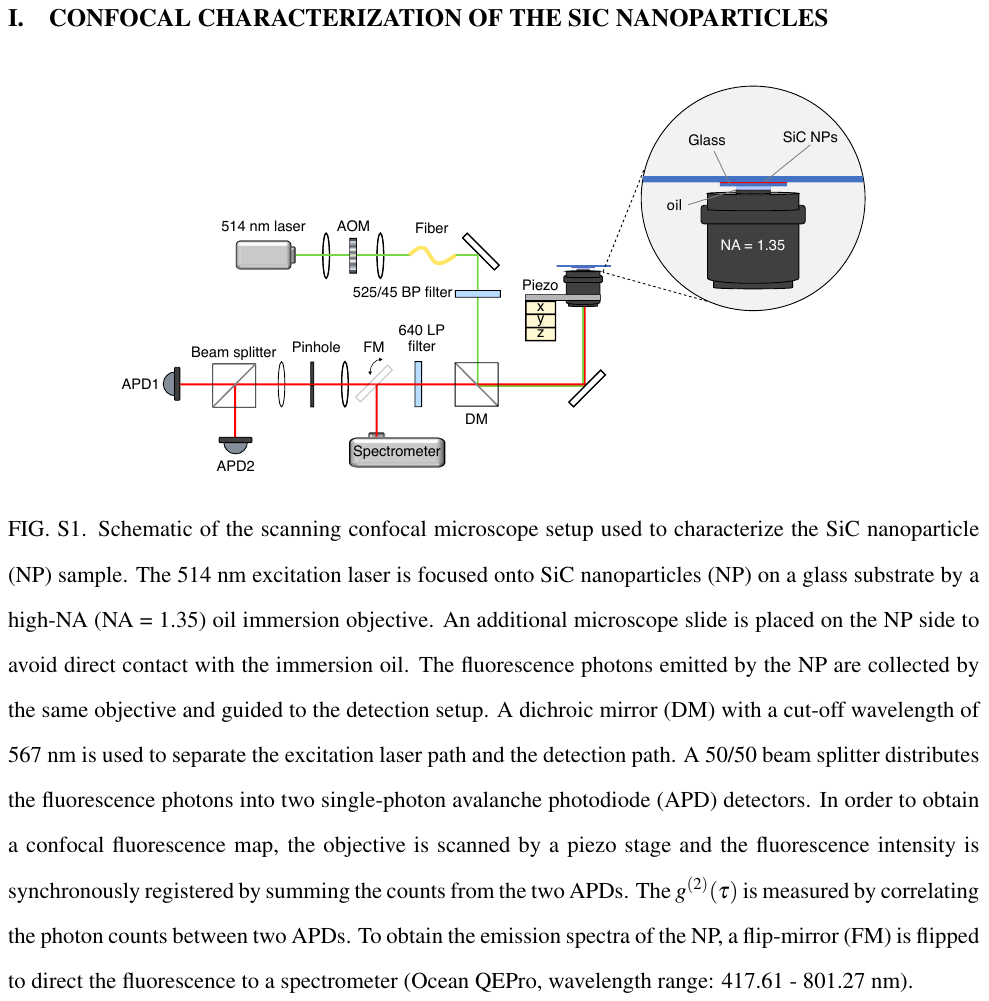}
   \end{figure*}

\begin{figure*}
\vspace*{-1cm}
\hspace*{-2cm}
    \includegraphics[page = 3, scale =1]{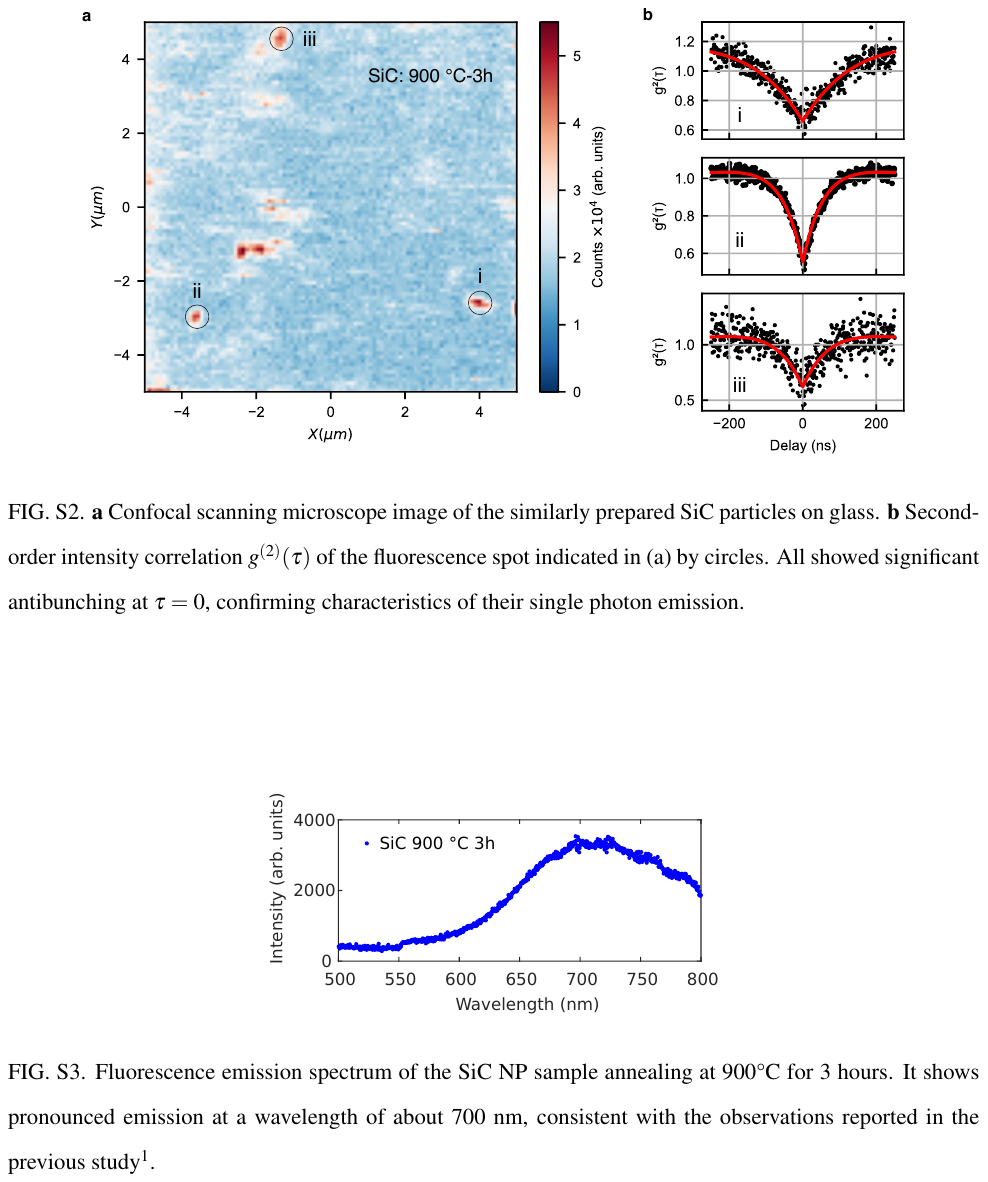}
   \end{figure*}

\begin{figure*}
\vspace*{-2cm}
\hspace*{-2cm}
    \includegraphics[page = 4, scale =1]{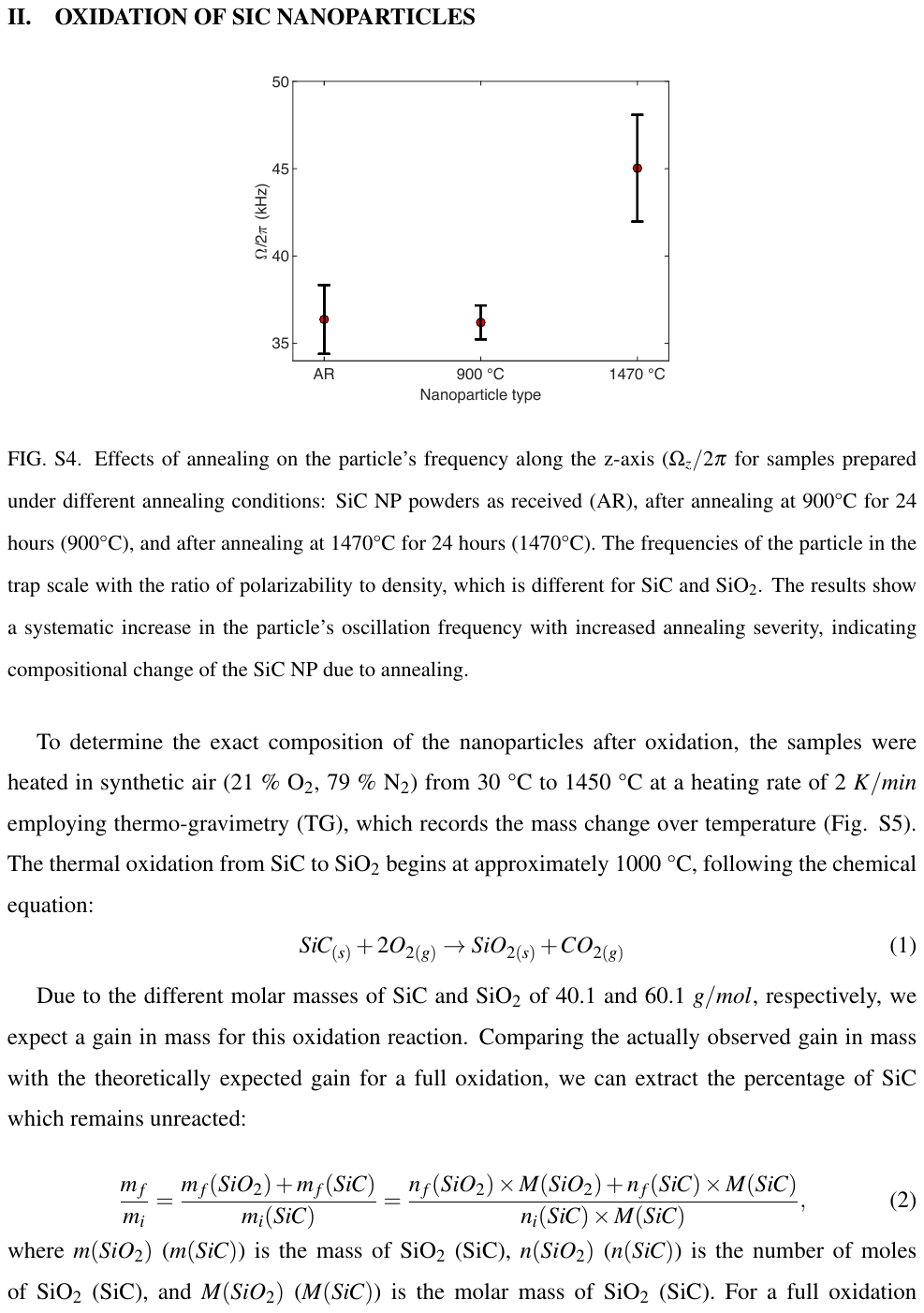}
   \end{figure*}

\begin{figure*}
\vspace*{-1cm}
\hspace*{-2cm}
    \includegraphics[page = 5, scale =1]{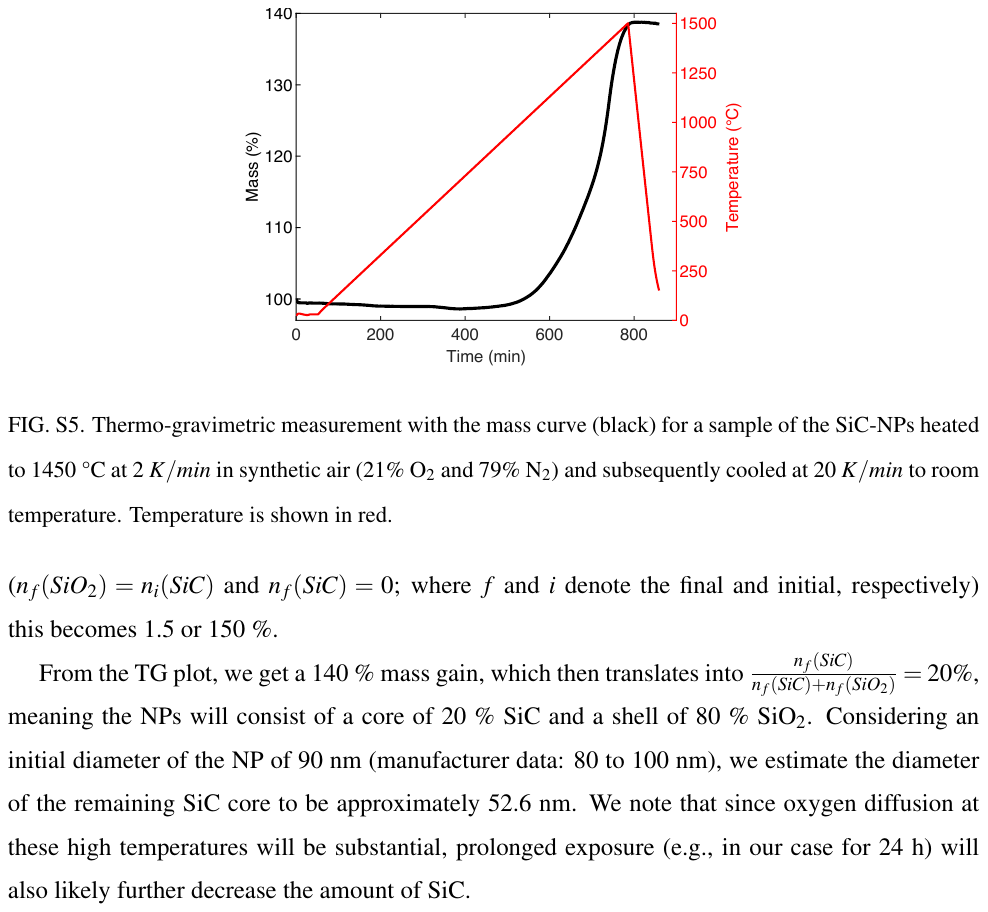}
   \end{figure*}

\begin{figure*}
\hspace*{-2cm}
    \includegraphics[page = 6, scale =1]{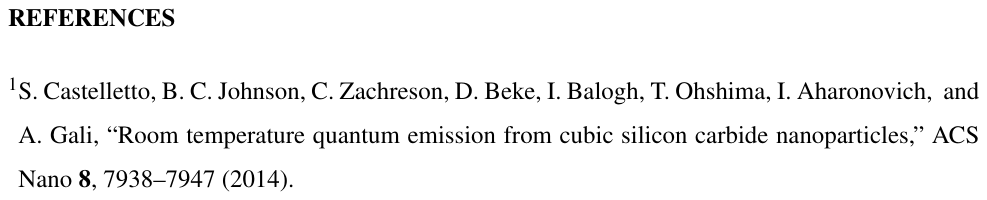}
   \end{figure*}
   
\end{document}